\documentstyle[russian,amsfonts,amssymb]{article}

\begin{document}
\begin{center}
{\bf Electrodynamics of accelerated charges or Why electron does not
radiate in Rutherford's atom}\\

    Stanislav Tereshchenko\\
\end{center}

\begin{abstract}
It is shown, that the radiation of the charge, moving with uniform
acceleration or uniformly moving round a circle and also freely
moving in a gravitational field, contradicts the principle of
equivalence. It is also shown, that the interaction of the charges,
moving with uniform acceleration or uniformly circling, which has
been calculated within the framework of classical electrodynamics,
leads to the violation of laws of conservation of energy, impulse
and angular momentum. We have offered a method in which way to
conform electrodynamics to the principle of equivalence. So in the
electrodynamics, which has been conformed in such a way, all the
mentioned violations of the laws of conservation are automatically
removed and the stability of Rutherford's atom is explained. It is
shown that the changes, which we have brought into the
electrodynamics, do not contradict the results of experiments.
\end{abstract}

\section{The violation of the principle of equivalence}

The principle of equivalence which is the base of the general
theory of relativity reflects the close connection between inertial
coordinate system $K$ with uniform gravitational field and
noninertial coordinate system $K^{\rm \prime}$ moving with
uniform acceleration in empty space. According to this principle any
physical process proceeds absolutely identically in the both
systems. In other words, if we imagine the system $K$ as a closed
laboratory, which is at rest on the Earth, and the system $K^{\rm
\prime}$ as an identical laboratory, moving with acceleration g in
a distance of gravitating masses, and the sizes of the laboratories
are such chosen, that it will be possible to neglect the
nonuniformity of the gravitational field in system $K,$ then the
observer, who is in one of these laboratories and who does not have
any connection with the external world, i.~e. without having any
possibility to look out of its limits, could not make any
experiment inside the laboratory in order to find out in which of
them he is, i.~e. he could not define the character of his motion.
That can be explained by the fact, that in system $K^{\rm \prime}$
during the uniform accelerating the field of inertial forces
appears. The action of this field does not differ from the action of
the gravitational field.

It is easy to make sure, that the mechanical processes will proceed
identically in both laboratories. In fact, free bodies in every
laboratory will move with acceleration $g$, identical pendulums will
oscillate with equal periods, etc. Thus, systems $K$ and $K^{\rm
\prime}$ are equivalent in respect of mechanical processes.

Einstein spread the equivalence of systems $K$ and $K^{\rm \prime}$
on all physical processes without exceptions, having formulated the
principle of equivalence, according to which, not only mechanical,
but also any physical processes have to proceed identically in
systems $K$ and $K^{\rm \prime}$.

But we can point to the process which violates the principle of
equivalence. It is the radiation of charges. It is easy to make
sure, that with the help of charges it is possible to differ system
$K$ from system $K^{\prime}$. In fact, let us place the charges in
both laboratories. The charge in the laboratory $K^{\prime}$ has to
radiate, so it is moving with acceleration. The observer who is in
this laboratory can register this radiation having placed, for
example, a charge into the water. Then a part of the radiating
energy will be absorbed by the water and will warm it up. On
measuring the temperature of the water the observer will be to
register the radiation.

We won't discuss the technical details how to carry out such an
experiment. It is enough that it is a principal possibility to find
out the radiation in immediate proximity to the charge.

In laboratory $K$ a motionless charge will not radiate. Thus, the
observer in every laboratory very easily can define the character
of his motion.

Therefore the principle of equivalence is violated.

Now we will consider uniformly rotating coordinate systems.

We can imagine one of these systems (we mark it as $S^{\prime}$) as
a disk, revolving with constant angular velocity on its axis. In
accordance with the principle of equivalence the noninertial
coordinate system $S^{\prime}$, in which a field of centrifugal
forces of inertia exists, can be considered as an inertial system
$K$ with an uniform gravitational field.

Let us consider two charges: one of them is on the disk, i.~e.
rotates with this disk. The second one is at rest at system $K$. In
the first case the charge has to radiate and in the second one it
does not. It has been already told how to differ a radiating charge
from a charge, which does not radiate. And what is more, a braking
force of radiation friction $f$ must act on uniformly moving round a
circle charge, which action one can observe at an as short as we
want distance from the charge.

Therefore, in this case the principle of equivalence is violated.

So far as we have mentioned the radiation force of friction, it is
necessary to point out another problem, which was first remarked by
M.~Born. By motion of the charge with uniform acceleration, force
$f$ turns into zero. That leads to the violation of energy balance.
Indeed $\displaystyle f=\frac{2e^{2}}{3c^{3}}\cdot
\mathop{r}_{}^{\ldots{}}$, where
$\displaystyle \mathop{r}_{}^{\ldots{}}$~--- the third derivative
from a coordinate.  If the motion firmly accelerates,
$\ddot{r}$=const and $\displaystyle \mathop{r}_{}^{\ldots{}}$=0.
Therefore, an uniformly accelerating charge radiates without any
losses of energy. That contradicts the law of conservation of
energy.

We will return to laboratories $K$ and $K^{\prime}$ again. We will
change the character of their motion. Let laboratory $K^{\prime}$,
which is in empty space far from gravitating masses, move uniformly
and straightforward and let $K$ freely move in a gravitational
field. There will be state of weightlessness in both laboratories.
According to the principle of equivalence, in this case systems $K$
and $K^{\prime}$ are also equivalent. I.~e. as in the last
instance, the observer, who is in one of the laboratories and does
not have any connection with the external world, could not make any
experiment inside the laboratory in order to find out the character
of his motion.

But we will place a charge in each of these laboratories again. In
laboratory $K$ charge must radiate, for it moves with acceleration
under the action of the gravitational field. The observer in this
laboratory will be able to register the radiation. In laboratory
$K^{\prime}$ the charge will not radiate.

Therefore, in this case the principle of equivalence is not
fulfilled.

\section{The violation of laws of conservation in classical
electrodynamics}

By means of not complicated calculations it can be shown, that
there is a violation of laws of conservation of energy, impulse and
angular momentum in classical electrodynamics. And the violation of
these laws arises when we consider the charges moving with uniform
acceleration or uniformly moving round a circle~--- i.~e. those
charges, the electrodynamics of which violates the principle of
equivalence. One can make sure in that, having calculated the
interaction of such charges.

Let charge $e$ move arbitrarily. Its field $\vec{E}$ is determined
with the following expression

$$\displaystyle
\vec{E}=\frac{e\left(1-\frac{v^{2}}{c^{2}}\right)}
{\left(R-\frac{\vec{R}\
\vec{v}}{c}\right)^{3}}\left(\vec{R}-\frac{\vec{v}}{c}R\right)
+\frac{e}{c^{2}\left(R-\frac{\vec{R}\,\vec{v}}{c}\right)^{3}}
\left[\vec{R}\left[\left(\vec{R}-\frac{\vec{v}}{c}R\right)
\displaystyle \mathop{\vec{v}}^{\cdot}\right]\right] \eqno(1)$$

\noindent
All the quantities in the right part are taken at the moment
$t^{\prime}$, which precedes the moment $t$ of the observation and
which can be obtained from the equation

$$\displaystyle t^{\prime}+\frac{R(t^{\prime})}{c}=t$$

\noindent
where $R(t^{\prime})$ is a distance from the charge to the point of
observation and $\vec{v}$ and $\displaystyle
\mathop{\vec{v}}^{\cdot}$ are velocity and acceleration of the
charge. The field consists of two parts. The first term describes
Coulomb's field of the charge, and the second one describes the
field of radiation.

We will consider two nonrelative charges $e$ and $-e$, connected
with a pivot, which length is $l$. These charges are moving with
acceleration $\displaystyle \mathop{\vec{v}}^{\cdot}$, directed
along the pivot (fig.~1). Let us define Coulomb's field
$\vec{E}_{k1}$, which the charge $e$ creates in the point $C$,
where charge $-e$ is. At the moment $t^{\prime}$ the charge $e$ is
in point $A$. Vector $\vec{R}(t^{\prime})$ is equal to the length
of segment $AC$ by its module. Vector $\displaystyle
\frac{\vec{v}}{c}R$ is equal to the length of segment $AB$
(fig.~1). Point $B$ is a position, in which charge $e$ could be at
the moment $t$, as if it were moving from point A uniformly with
velocity $v(t^{\prime})$ during the period of time
$\displaystyle t-t^{\prime}=\frac{R}{c}$. But the charge $e$ is
moving with acceleration $\displaystyle \mathop{\vec{v}}^{\cdot}$.
That is why at the moment $t$ it will be in point $D$. And segment
$\displaystyle AD=v\tau +\frac{\dot{v}\tau^{2}}{2}$, where $\tau
=t-t^{\prime}$.

Therefore the Coulomb's field $\vec{E}_{k1}$ is equal to

$$\displaystyle E_{k1}=\frac{e}{\left
(l-\frac{\dot{v}\tau^{2}}{2}\right)^{2}}\,.$$

\noindent
For $v\ll c$, then $R(t^{\prime})\approx l$, $\displaystyle \tau
\approx\frac{l}{c}$, then $\displaystyle
E_{k1}=\frac{e}{\left(l-\frac{\dot{v}l^{2}}{2c^{2}}\right)^{2}}\,.$


Therefore, force $\vec{F}_{1}$, acting on the charge $-e$, is equal
to

$$\displaystyle F_{1}=\frac{e^{2}}{\left(
l-\frac{\dot{v}l^{2}}{2c^{2}}\right)^{2}}\,.$$

In the same way we will determine field $\vec{E}_{k2}$, which the
charge $-e$ creates in point $D$, where the charge $e$ is. At the
moment $t^{\prime}$ the charge $-e$ will be in point $K$ (fig.~1).
The segment $KL=v\tau$. The segment $\displaystyle KC=v\tau
+\frac{\dot{v}\tau^{2}}{2}\,.$

\noindent
So field $\vec{E}_{k2}$ is equal to

$$\displaystyle E_{k2}=\frac{e}{\left(
l+\frac{\dot{v}l^{2}}{2c^{2}}\right)^{2}}\,.$$

\noindent
The force $\vec{F}_{2}$, acting on the charge $e$ is equal to

$$\displaystyle F_{2}=\frac{e^{2}}{\left(
l+\frac{\dot{v}l^{2}}{2c^{2}}\right)^{2}}\,.$$

Thus, forces $\vec{F}_{1}$ and $\vec{F}_{2}$ are not equal by the
quantily. Their resultant $\Delta \vec{F}=\vec{F}_{1}+\vec{F}_{2}$
is

$$\displaystyle \Delta F=\frac{2e^{2}\dot{v}}{c^{2}l}$$

\noindent
and is directed along the vector $\displaystyle
\mathop{\vec{v}}^{\cdot}$.

\noindent
The forces of the interaction of the
charges, which are specified by the second item in (1), are equal
to zero in this case.

In the case, when vector $\displaystyle \mathop{\vec{v}}^{\cdot}$
and the axis of the pivot make angle $\alpha$, the forces
$\vec{F}_{1}$ u $\vec{F}_{2}$ are not equal by the quantity and
do not lie on the same straight line (vectors $\vec{R}$ and
$\displaystyle \frac{\vec{v}}{c}R$ of the charge $e$ are shown in
fig.~2).

\noindent
The projection of their resultant $\Delta\vec{F}$ on axis $x$ is
equal to

$$\displaystyle \Delta F_{x}=\frac{e^{2}\dot{v}}{c^{2}l}(2\cos
^{2}\alpha -\sin ^{2}\alpha)$$

\noindent
and on axis $y$

$$\displaystyle \Delta F_{y}=-\frac{3e^{2}\dot{v}}{2c^{2}l}\sin
2\alpha.$$

The forces of the interaction of the charges, specified by the
second item in (1), are not equal to zero in the ammount in this
case either. The projection of its resultant $\Delta F^{\prime}$ on
axis $x$ and axis $y$ are equal

$$\displaystyle \Delta F^{\prime}_{x}=\frac{2e^{2}\dot{v}}{c^{2}l}
\sin ^{2}\alpha;$$

$$\displaystyle \Delta F^{\prime}_{y}=\frac{e^{2}\dot{v}}{c^{2}l}
\sin 2\alpha.$$

It is quite evidently, that the presence of the resultant $\Delta
\vec{F}$, depending on the orientation of the pivot and acting on
the electrically neutral system, contradicts the laws of
conservation of energy and impulse. Within the framework of
classical electrodynamics it is impossible to compensate this
force.

We will consider the following instance. Let cylinder of radius $r$
the forming $d$ revolve on its axis. We should choose quantaties
$r$ and  $d$, so as 2$\pi r=0,01$~$d$. At the ends of the forming we
place charges $e$ and $-e$ (fig.~3). Let linear velocity of the
charges be equal to $v=0,01~c$. In order to calculate Coulomb's
field $\vec{E}_{k}$, which the charge $e$ creates in point $A$,
where charge $-e$ is, it is necessary to find out the preceding
position of charge $e$ at moment $t'$~--- as in the previous
instance. It is evidently, that for this case the period of time
$t-t'$ is equal to the period of revolution of the cylinder. So, at
the moment $t'$ the charge $e$ will be in point B, where it is also
at the moment $t$. Vectors $\displaystyle \vec{R},
\,\frac{\vec{v}}{c}R$ and $\displaystyle \vec{R} -
\frac{\vec{v}}{c}R$ will be directed as it is shown in fig.~3\,
($BP=2\pi r$). The field $\vec{E}_{k}$ in point $A$ will
be equal $E_{k}\approx\displaystyle \frac{e}{d^{2}}$ and will
direct along the segment $PA$.

That is why Coulomb's force, acting on the charge $-e$, has got
the component on the tangent to the point $A$. The quantity of this
component is equal to $F=0,01$ F$_{k}$ where $F_{k}$ is
Coulomb's force.

The force, acting on the charge $-e$, specified by the second
item in(1), will be directed perpendicularly to the velocity of the
charge $-e$.

In the same way we will calculate the effect of charge $-e$ on
charge $e$. There fore on the cylinder under review, which is a
closed system, the rotational moment $M=0,02$ F$_{k}r$ is
acting. That contradicts the principles of conservation of energy
and angular momentum.

It is easy to show, that if the velocity of the rotation of this
cylinder is diminished twice, Coulomb's interaction of charges $e$
and $-e$ will lead to the braking of the cylinder.

\section{Rutherford's atom}

One of the most important problems of classical electrodynamics is
the problem of the stability of Rutherford's atom. As it is known,
the planetary model of atom, offered by Rutherford, has a principal
shortcoming~--- it was unstable. According to classical
electrodynamics, an electron, moving round a circular orbit, has to
radiate and as a resultat it will fall down  on the nucleus. Bohr's
postulates, which had appeared soon after that, did not make clear
this problem~--- the existance of allowed orbits also contradicted
classical electrodynamics.

The quantum mechanics does not explain either, why electron in a
stationary orbit does not radiate. The radiation of the electron,
when it is going from one allowed orbit to another one, is not
connected with the radiation of an accelerated electron anyhow.
Atom can be in an excited state rather long, but an accelerated
electron must radiate constantly. And what is more, in quantum
mechanics the concept of movement with acceleration is not
considered at all. The reference to in applicability of classical
electrodynamics to atomic objects is quite unfounded. In fact,
electron is held by forces, which submit to principles of
electrodynamics. Then, why don't these principles spread on motion
of electrons?

Thus, the problem of the stability of atom remains to be unsolved
up to now.

\section{Do inertial coordinate systems exist?}

In classical electrodynamics motion of charges is considered in
inertial coordinate systems. These systems are defined in the
following way: if any forces do not act on the body, or the sum of
acting forces is equal to zero, then there are such coordinate
systems, relative to which the body is moving without any
acceleration. These systems are named inertial (The first Newton's
law). Inertial systems are moving uniformly and straigtforward
relatively to each other, so the acceleration of the body has
identical value in any of these systems, i.~e. acceleration is
absolute quantity.

But how can we realise inertial coordinate system in practice?
Obviously, in order to realise it, it is necessary to take a body,
on which any force does not act at all, or their sum must be equal
to zero, and then to find a coordinate system, relatively to which
this body will be moving without any acceleration. That system will
be inertial.

But it is impossible to isolate a body from any action of external
forces. Such isolation would mean, that there is only one body in
the Universe. Thus, we can only try to bring the sum of external
actions to zero. But how can we find out, that the sum of forces,
acting on a body, is equal to zero? We can do it only in the
following way~--- to define the acceleration on the body relatively
to inertial coordinate system. If the acceleration is equal to
zero, then the sum of the forces is equal to zero.

So we have got an exclusive circle; in order to determine an
inertial coordinate system we have to choose a body so as the sum
of acting forces on it is equal to zero; but we need to have an
inertial coordinate system in order to determine equality of this
sum of external forces to zero.

Thus, we have got a real difficulty when we try to realize the
inertial coordinate system in practice.

In principle there is a possibility in classical electrodynamics to
realize the inertial coordinate system. In fact, according to
electrodynamics an accelerated charge must radiate. The intensity
of radiation (in nonrelativistic case) is defined with an
expression
$$\displaystyle I=\frac{2e^{2}w^{2}}{3c^{3}}          \eqno (2)$$
\noindent
where {\it e} is a quantity of the charge, and {\it w} is its
acceleration. Equations of electrodynamisc are true in inertial
coordinate systems.

That is why, if a charge moves with acceleration, it radiates, if
its acceleration is equal to zero, there is no any radiation.
Radiation is an absolute quantity. It is impossible to create or to
destroy it by any choose of any coordinate system. So the
acceleration, which is undoubtedly connected with radiation, has to
be an absolute quantity. Therefore, if some charge does not
radiate, the system, which is connected with it, will be inertial.

In classical physics it is supposed, that equations of
electrodynamics are true only in inertial coordinate systems. It is
easy to make sure, that such a point of view is not founded enough.
Indeed, the equations of Maxwell were received as a result of the
generalization of experimentals. So far as there already were
distinguished inertial coordinate systems, the equations of Maxwell
were brought to these systems. But we do not have any grounds to
affirm, that the experimentals, in which electromagnetic phenomens
are studied and which are carried out in inertial coordinate
systems and the results of the same experiments, which are carried
out in an uniformly accelerated coordinate system, will be
different. And if the results of these experiments in these systems
are identical, in will mean, that the equations of Maxwell will be
also true in uniformly accelerated systems.

Will Coulomb's law be true in uniformly accelerated systems? In
other words, if one can measure the interaction of motionless
charges in a laboratory, which moves with uniform acceleration,
then can this interaction be described by Coulomb's law? In
particular, will the forces of the interaction of the charges be on
the same straight line? Obviously, that we can answer this question
only on the grounds of the experimental. According to classical
physics, Coulomb's law is not true in uniformly accelerated
coordinate system, for the charges, which are at rest in this
system, move with acceleration in inertial coordinate systems, and
the electromagnetic field of such charges is described by means of
lagging potentials. That is why, Coulomb's forces, with which the
charges, which are at rest in the accelerated laboratory, interact,
do not lie on the same straight line. But in the first place, this
conclusion is not confirmed with experimentals, and in the second
place, as we have seen, Coulomb's interaction of the charges,
moving with uniform acceleration and calculated within framework of
classical electrodynamics leads to the violation of laws of
conservation of energy and impulse.

So the fact, that there is not any radiation of the charge, is not
a reason to suppose, that the coordinate system, which is connected
with the charge, is inertial. That fact only indicates, that all
electromagnetic processes in this system will proceed in the same
way as in the system, which is connected with the Earth.

But if we are not able to realize the inertial coordinate system,
then all the systems become equal. In this case the acceleration
loses its absolute sense and becomes a quantity, which is as
relative as the velocity. So as to determine the intensity of the
radiation, the quantity of the acceleration in expression (2) has
to be chosen so as to take additional terms into account.

It is remarkable, that the concept of the equality of all
coordinate systems has been completely realised in general theory
of relativity. And what is more, if we managed to realise the
inertial coordinate system, the general theory of relativity
would turn out groundless.

\section{The electrodynamics and the Principle of equivalence}

We will try to remove the contradictions, which have been noticed
before. First we will consider a charge moving freely in a
gravitational field. So as the principle of equivalence will be
true, this charge must not radiate. This conclusion seems to be
rather logical. Indeed, it is supposed in classical physics, that
there is a straight line, transpiercing through all the Universe
and a charge, which is moving along it with constant velocity and
does not radiate. Any deflection of such movement must be
accompanied by the radiation. In classical physics only a ray of
light must be a straight line. But light declines in a
gravitational field. So one can get a straight line only when there
is no any gravitational fields, i. e. in a limited spheres of
space.

In the general theory of relativity a concept of geodesic line is
introduced. It is a generalized notion of a straight line. It is a
trajectory of the motion of a body, on which any forces do not act,
except the gravitational ones. Such movement of a body in the
general theory of relativity is inertial. So we can expect, that
the free movement of a charge in the gravitational field will not
be accompanied by radiation. That means, that in the coordinate
system, freely moving in a gravitational field, the equations of
Maxwell will have the same form, as in the inertial system.

Let us consider the charge $q$, which is at rest in the inertial
coordinate system with a constant gravitational field. Evidently,
vector $\displaystyle \mathop{E}^{\to}$ of such charge in any
coordinate system will lie on a straight line, connected the charge
and the point of observation. Indeed, let a charge $e$ lie at a
distance of charge $q$. The force, acting on the charge $e$ from
the side of the charge $q$, will be on a straight line, connected
these charges. It is quite evidently, that the observer can move in
any way, i. e. he can be in any coordinate system, but the
direction of the force, acting on the charge $q$, will not change.
Therefore field $\displaystyle \mathop{E}^{\to}$ of the charge $q$
has to be described with expression

$$\displaystyle \mathop{E}^{\to}=
\frac{q \displaystyle \mathop{R}^{\to}}{R^{3}}\cdot \frac{1-
\frac{v^{2}}{c^{2}}}{(1- \frac{v^{2}}{c^{2}}{\rm sin}^{2}
\theta)^{3/2}}, \eqno (3)$$

\noindent
which is to be true for this charge in any coordinate system. Here
$R$ is a distance from charge $q$ to the point of the observation,
$\theta$ is an angle between the direction of the motion
of the charge and radius-vector
$\displaystyle \mathop{R}^{\to}$, $v$ is
momentary velocity of the charge relative to the observer.
Expression (3) describes the field of a uniformly moving charge.
Its applicability in this case evident. Indeed, let there be an
accelerated system and an inertial system, accompanying it. It is
clear, that field $\displaystyle \mathop{E}^{\to}$ in the both
systems will be identical at any moment for the observers.

For the charge $q$, which is at rest in a constant gravitational
field, does not radiate then so as to fulfil the principle of
equivalence, the charges, moving with uniform acceleration or
uniformly moving round a circle, must not radiate either. Field
$\displaystyle \mathop{E}^{\to}$ of such charges must also be
described with expression (3), which is true for these charges in
any coordinate system. (We consider only those charges, the
quantity of the acceleration of which does not change during the
period of observation, i. e. we except transitional processes).
That means, that in the coordinate systems, which moves with an
uniform acceleration or uniformly moves round a circle, Maxwell's
equations has the same form as in the inertial systems. Under this
assertion we mean the following: in the systems, moving with an
uniform acceleration or uniformly moving round a circle, all
electromagnetic processes will proceed in the same way as in the
inertial systems with constant gravitational field.

But if the field $\displaystyle \mathop{E}^{\to}$ of the charges,
moving with uniform acceleration or uniformly moving round a
circle, is described with expression (3), there is no any violation
of the laws of conservation by calculation of its interaction.
Indeed, in this case the forces, which the charges, which were
considered before, interact with, will be equal by the quantity and
\ will \ be \ contrary \ directed.

The lack of the force of radiation friction during the motion with
uniform acceleration becomes clear. For such charge does not
radiate, so there will not be any violation of the balanse of
energy.

How do the changes, which have been brought into electrodynamics,
accord with the experimentals? It is the first question. And the
second one~--- if the charge radiates, what value of the
acceleration should we put down into the expression(2) to define
intensity of the radiation.

Let us answer the second question first. We will consider a charge,
oscillating in accordance with harmonic law relatively to inertial
system $K$. Under inertial system $K$ we mean here coordinate
system, connected with the Earth. By such movement of the charge
contradictions do not appear. So the charges, the motion of which
can be brought to harmonic oscillations in system $K$, we will
describe within the framework of classical electrodynamics. Now let
the charge move straightforward with acceleration $w(t)$ in system
$K$. If function $w (t)$ is periodic, we can expand it in
trigonometric row

$$w(t)=w_{0}+ \mathop{\Sigma}_{n=1}^{\infty}(b_{n}\cos
\omega nt+c_{n} \sin \omega nt), \eqno (4)$$

\noindent
where $w_{0}$~--- constant of function $w(t)$, which is accorded
with the movement of the charge with uniform acceleration and
$b_{n}$ and $c_{n}$~--- amplitudes of harmonic $n$. I. e.
the function $w(t)$ can be expanded in a form of a sum of two items
$w(t)=w_{0}+w_{1}(t)$. Through $w_{1}(t)$ the sum of harmonics is
designated in (4). For we think, that the uniformly accelerated
charge does not radiate, then in the expression (2) for intensity of
the radiation only value $w_{1}(t)$ has to be put in as an
acceleration. If function $w(t)$ is not periodic, it has to be
expanded into Fourier's integral. In this case the steady component
$w_{0}$ is equal to zero. So we will put the complete value of the
acceleration $w(t)$ into expression (2), i. e. the radiation of the
charge will be defined in the same way as in classical dynamics.
For in practice a charge can move straightforward only during a
short period of time, its acceleration cannot be a periodic
function. The radiation of such a charge therefore will completely
correspond to parameters, which have been received within the
framework of electrodynamics.

During the movement of a charge in a closed trajectory with a
constant period of revolution, the value of its acceleration $w(t)$
can be also expanded in a trigonometric row (4). The steady
component $w_{0}$ corresponds to the acceleration of the uniform
moving round a circle here. So when we calculate the intensity of
the radiation in expression (2), as an acceleration we will put
in only the value $w_{1}(t)$. If function $w(t)$ is not periodic, we
will put in the complete value of the acceleration.

\noindent
The fact, that the charge, uniformly moving  round a circle, does
not have any radiation, does not contradict the existence of the
synchrotron radiation. Indeed, the quantity of the acceleration of
electrons in cyclical accelerators is not a periodic function. If
we consider the movement of electron in the accelerator as
periodic, we will receive the following resultat. The intensity of
radiation I$_{ 1}$ on frequency of circulation $w_{1}$ (the
first harmonic), calculated within the framework of classical
electrodynamics, is proportionate to
$$\displaystyle w_{0}^{2}+\frac{1}{2}(b_{1}^{2}+c^{2}_{1}).      $$
The intensity of the $nth$ harmonic is proportionate to
$\displaystyle \frac{1}{2}(b^{2}_{ n}+c^{2}_{ n}).$ If we
suppose, that the charge, uniformly moving round a circle, does not
radiate, it will only bring us to the fact, that only the intensity
of the first harmonic will decrease. In this case it must by
proportionate to $\displaystyle \frac{1}{2}(b^{2}_{1}+c^{2}_{1}$).
The intensity of the rest harmonics will not change. As it is known
from the electrodynamics, the great part of the radiation is
concentrated in the range of frequencies $\displaystyle w \sim
w_{1}\left(\frac{\varepsilon}{mc^{2}}\right)^{3}$, where
$\varepsilon$~--- is an energy of electron. On frequency $w_{1}$ a
small part of energy is radiated. So when
$\varepsilon =50$~Mev \,$I_{1}/I \sim 10^{ -8}$, where $I$ is a
complete intensity. It is necessary to take notice of the fact,
that because of the influence of the conductive surfaces (the sides
of the vacuum chamber of an accelerator) the intensity of the low
frequency part of the radiation will decrease approximately
$\displaystyle \left(\frac{r}{d}\right)^{2}$ times, where $r$ is a
radius of the orbit, and $d$ is a distance from the electron beam
to the conductive surface.

So it is clear, that it is practically impossible to notice such
deflection in synchrotron radiation. We need a special experiment,
which could help us to test the intensity of the radiation on
frequency of revolution. When the charge uniformly moves round a
circle, there must not be any radiation at all. In such way
electron in atom can move in an stationary orbit. And as it is
known, such electron does not radiate. This fact confirms the
results we have received. On the other hand, we have received the
explanations of the stability of Rutherford's atom, i. e. why
electron does not radiate in a circular orbit.

\section{Supplement}

As it is known, the equations of physics, which are written in
general covariance form, automatically satisfy the principle of
equivalence. Therefore the general covariance expression for the
intensity is not to violate the principle of equivalence. Let us
receive such an expression. The charge is at rest in the inertial
coordinate system. Its energy, which was radiated during period
$dt$, is equal to
$$\displaystyle d\varepsilon=
   \frac{2e^{2}w^{2}}{3c^{2}}dt.
 \eqno (5)$$

\noindent
Complete radiated impulse in this coordinate system is equal to
zero
$$d\vec{P}=0.    \eqno (6)$$

\noindent
In order to pass to an arbitary inertial coordinate system we will
write down expressions (5) and (6) in four-dimensional form
$$\displaystyle dP^{ i}=-\frac{2e^{2}}{3c}\,\frac{du^{
k}}{ds}\cdot\frac{du_{ k}}{ds}u^{i}ds.          \eqno (7)$$

\noindent
So as to pass to general covariance expression for $dP^{i}$ in (7)
we will substitute usual differentiation for covariance. Then we
will receive
$$\displaystyle dP^{
i}=-\frac{2e^{2}}{3c}\left(\frac{du^{ k}}{ds}+\Gamma_{
l\,m}^{ k}u^{ l}u^{ m}\right)\,\left(\frac{du_{
k}}{ds}-\Gamma ^{ i}_{ kl}u_{ i}u^{ m}\right)u^{
i}ds.                                              \eqno (8)$$

\noindent
Here $\Gamma ^{ k}_{ l\,m}$ are symbols of Christoffel, which
are
$$\displaystyle \Gamma^{k}_{ i\,l}=\frac{1}{2}g^{
k\,n}\left(\frac{\partial g_{ n\,l}}{\partial x^{ m}}+
\frac{\partial g_{ n\,m}}{\partial x^{ l}}-\frac{\partial
g_{ l\,m}}{\partial x^{ n}}\right),                  \eqno (9)$$

\noindent
where $g^{ l\,m}$ is a fundamental tensor. Indexes take values
0, 1, 2, 3. Expression (8) is true in any coordinate system. For a
body, moving freely in the gravitational field, it is
$$\displaystyle \frac{du^{ k}}{ds}+\Gamma ^{ k}_{
l\,m}u^{l}u^{ m}=0.                   $$

\noindent
Therefore, the charge, moving in such way, does not radiate. We
have received this conclusion before, when we were according
electrodynamics with the principle of equivalence.

Now we will consider the charge, which is motionless in inertial
coordinate system $K$ with constant gravitational field. The
acceleration of such charge in system $K$ is zero, and the
components of the 4-velocity are
$$u_{0}=1, \,u_{1}=u_{2}=u_{3}=0.                                 $$

\noindent
So it follows from (8), that the intensity of radiation of the
charge in this case is
$$\displaystyle
\frac{d\varepsilon}{dt}=\frac{2e^{2}c}{3}\Gamma^{ i}_{
0\,0}\Gamma^{0}_{i\,0}.                                         $$

\noindent
For a weak gravitational field components of tensor $g_{e\,m}$ are
$$\displaystyle g_{1\,1}\approx g_{2\,2}\approx g_{3\,3}\approx -1,
\, g_{0\,0}=1+\frac{2\varphi}{c^{2}}, \,g_{ i\,k}\approx 0
\,(i\not=k),                                                         $$

\noindent
where $\varphi$ is a gravitational potential. Taking the statical
nature of the gravitational field into consideration, i.~e.
$\displaystyle \frac{\partial g_{k\,l}}{\partial x^{0}}=0$, and
that the components of tensor $g^{ik}$ in this case have
approximately the same values, as the components of tensor
$g_{i\,k}$, we will receive, when $\alpha$=1, 2, 3.
$$\displaystyle \Gamma^{\alpha}_{00}=\Gamma^{0}_{\alpha
0}=\frac{1}{2}\,\frac{\partial g_{00}}{\partial x^{\alpha}}.   $$

\noindent
Therefore, the intensity of radiation is equal to
$$\displaystyle
\frac{d\varepsilon}{dt}=\frac{2e^{2}}{3c^{3}}\cdot\frac{\partial
\varphi}{\partial x^{\alpha}}\cdot \frac{\partial \varphi}{\partial
x_{\alpha}}.                                          \eqno(9)
$$

\noindent
For $\dot{\vec{v}}=-\nabla \varphi$, and acceleration
$\dot{\vec{v}}$ in this case is the accelaration due to gravity $g$,
then (9) will have the following form $$\displaystyle
\frac{d\varepsilon}{dt}=\frac{2e^{2}}{3c^{3}}g^{2}.  $$

\noindent
Thus, the charge, which is immovable on the Earth, must radiate.
The intensity of its radiation is identical to the intensity of the
charge, moving with acceleration $g$ in empty space at a great
distance away from any gravitating bodies. So the principle of
equivalence is formally fulfilled. It is naturally for covariance
expression (8), though it is clear, the charge, which is at rest in
a gravitational field, does not charge.

\end{document}